# A Reconsideration of Matter Waves

by

*Roger Ellman*

Abstract

Matter waves were discovered in the early 20th century from their wavelength, predicted by DeBroglie, Planck's constant divided by the particle's momentum, that is, $\lambda_{mw} = h/m \cdot v$. But, the failure to obtain a reasonable theory for the matter wave frequency resulted somewhat in loss of further interest.

It was expected that the frequency of the matter wave should correspond to the particle kinetic energy, that is, $f_{mw} = \frac{1}{2} \cdot m \cdot v^2 / h$ but the resulting velocity of the matter of the particle, $v = f_{mw} \cdot \lambda_{mw}$, is that the matter wave moves at one half the speed of the particle, obviously absurd as the particle and its wave must move together.

If relativistic mass is used (as it should in any case) the problem remains, the same mass appearing in numerator and denominator and canceling. It is no help to hypothesize that the total energy, not just the kinetic energy, yields the matter wave. That attributes a matter wave to a particle at rest. It also gives the resulting velocity as $c^2/v$, the wave racing ahead of its particle.

A reinterpretation of Einstein's derivation of relativistic kinetic energy (which produced his famous $E = m \cdot c^2$) leads to a valid matter wave frequency and a new understanding of particle kinetics and of the atom's stable orbits.

Roger Ellman, The-Origin Foundation, Inc.
    320 Gemma Circle, Santa Rosa, CA 95404, USA
    RogerEllman@The-Origin.org
    http://www.The-Origin.org

# A Reconsideration of Matter Waves

## by

## Roger Ellman

### The Matter Wave Problem

In the early 20th Century (1924) DeBroglie proposed that, since light, which was then considered to be a purely wave phenomenon, had been found to appear sometimes to exhibit particle behavior; perhaps matter, which was accepted as being particle in nature might sometimes exhibit wave behavior. DeBroglie reasoned that, the wavelength of a photon being equal to Planck's constant divided by the photon's momentum, the same relationship should apply to a particle of matter -- it having a wavelength of $h$ divided by the particle momentum.

The reasoning was as follows. First considering a photon, its energy is

(1) $\quad W_{wave} = h \cdot f$

and the energy equivalent of a mass, $m$, is

(2) $\quad W_{mass} = m \cdot c^2$

While the photon's rest mass is zero it has kinetic mass corresponding to its energy. If the photon equivalent mass, m, actually appears as a wave its energy as a wave must be the same as its energy as a mass. Therefore

(3) $\quad W_{mass} = W_{wave}$

$\quad m \cdot c^2 = h \cdot f$

$\quad m = \dfrac{h \cdot f}{c^2} \qquad$ [solving the above for m]

$\quad \phantom{m} = \dfrac{h}{\lambda \cdot c} \qquad$ [substituting $c = \lambda \cdot f$ for one of the c's in the denominator]

and, finally,

(4) $\quad \lambda = \dfrac{h}{m \cdot c} \qquad$ [solving (3) for λ]

$\quad \phantom{\lambda} = \dfrac{h}{\text{photon momentum}}$

recognizing that momentum is defined as the product of mass and its velocity and the velocity of the photon is $c$.

DeBroglie hypothesized that the wave aspect of a particle of matter should have an analogous wavelength, $\lambda_{mw}$, that should be

(5) $$\lambda_{mw} = \frac{h}{\text{particle momentum}} = \frac{h}{m \cdot v}$$

This suggestion of DeBroglie was soon verified by Davison and Germer who obtained electron diffraction patterns and found that the observed wavelengths of the electron matter waves corresponded well with DeBroglie's formulation.

At that point one would think that the duality of matter, as of light, was established and that extensive further investigation of matter waves would have resulted. But that was not the case and the reason was a fundamental problem that could not be overcome -- the matter wave frequency.

If one reasons, analogously to the derivation of $\lambda_{mw}$, that the kinetic energy of the particle of matter should correspond to its matter wave frequency, $f_{mw}$, as

(6) $$f_{mw} = \frac{W_k}{h} = \frac{\frac{1}{2} \cdot m \cdot v^2}{h}$$

then the velocity of the matter wave is

(7) $$v_{mw} = \lambda_{mw} \cdot f_{mw} = \left[\frac{h}{m \cdot v}\right] \cdot \left[\frac{\frac{1}{2} \cdot m \cdot v^2}{h}\right] = \frac{1}{2} \cdot v$$

a result that states that the matter wave moves at one half the speed of the particle. That is obviously absurd as they must move together each being merely an alternative aspect of the same real entity. For them not to move together would be as absurd as for the particle aspect of light to move at a different speed than its wave aspect, the photon not arriving coincident with the $E-M$ wave.

It is no help in resolving this difficulty if relativistic mass is used (as it should be in any case) since the same mass appears in both numerator and denominator of equation (7) where they simply cancel out.

It is also no help to hypothesize that it is the total energy, not just the kinetic energy, that yields the matter wave. Such an attempt attributes a matter wave to a particle at rest. It also gives the resulting matter wave velocity as $c^2/v$ which has the matter wave racing ahead of its particle. No, the two must keep pace with each other since they are the same thing merely looked at in one or the other of two alternative ways.

It was the inability to resolve this problem that led to the loss of interest in matter waves and essentially the end of further inquiry with regard to the wave aspect of matter.

### *Einstein's Derivation of Relativistic Kinetic Energy*

Kinetic energy, $KE$, is defined as the work done by the force, $f$, acting on the particle or object of mass, $m$, over the distance that the force acts, $s$. This quantity is calculated by integrating the action over differential distances.

$(8)$
$$KE = \int_0^s f \cdot ds \qquad \text{[Per above definition]}$$

$$= \int_0^s \frac{d(m \cdot v)}{dt} \cdot ds \qquad \text{[Newton's 2}^{nd}\text{ law]}$$

$$= \int_0^{(m \cdot v)} \frac{ds}{dt} \cdot d(m \cdot v) \qquad \text{[Rearrangement of form]}$$

$$= \int_0^{(m \cdot v)} v \cdot d(m \cdot v) \qquad [v = {ds}/{dt}]$$

$$= \int_0^v v \cdot d\left[\frac{m_r \cdot v}{\left[1 - \frac{v^2}{c^2}\right]^{\frac{1}{2}}}\right] \qquad \text{[m is } m_r \text{ Lorentz contracted by v. } m_r \text{ is rest mass]}$$

$$= \frac{m_r \cdot v^2}{\left[1 - \frac{v^2}{c^2}\right]^{\frac{1}{2}}} - m_r \cdot \int_0^v \frac{v \cdot dv}{\left[1 - \frac{v^2}{c^2}\right]^{\frac{1}{2}}} \qquad \text{[Integration by parts]}$$

$(9)$
$$KE = \frac{m_r \cdot v^2}{\left[1 - \frac{v^2}{c^2}\right]^{\frac{1}{2}}} - m_r \cdot c^2 \cdot \left[1 - \frac{v^2}{c^2}\right]^{\frac{1}{2}} - m_r \cdot c^2 \qquad \text{[Integration of 2nd term]}$$

$(10)$
$$= \frac{m_r \cdot v^2 + m_r \cdot c^2 \cdot \left[1 - \frac{v^2}{c^2}\right]}{\left[1 - \frac{v^2}{c^2}\right]^{\frac{1}{2}}} - m_r \cdot c^2 \qquad \text{[Place 2}^{nd}\text{ term over 1}^{st}\text{ term denominator]}$$

$$= \frac{m_r \cdot v^2 + m_r \cdot c^2 - m_r \cdot v^2}{\left[1 - \frac{v^2}{c^2}\right]^{\frac{1}{2}}} - m_r \cdot c^2 \qquad \text{[Expand term with brackets]}$$

$$= \frac{m_r \cdot c^2}{\left[1 - \frac{v^2}{c^2}\right]^{\frac{1}{2}}} - m_r \cdot c^2 \qquad \text{[Simplify]}$$

*(11)* $\quad KE = m_v \cdot c^2 - m_r \cdot c^2 \qquad$ [$m_v$ is total mass at $v \neq 0$
$\qquad\qquad\qquad\qquad\qquad\qquad\qquad\qquad m_r$ is total mass at $v = 0$
$\qquad\qquad\qquad\qquad\qquad\qquad\qquad\qquad m_v = m_r$ Lorentz transformed]

This result states that:

   {Kinetic Energy} = {Total Energy} - {Rest Energy}

or

   {Total Energy} = {Kinetic Energy} + {Rest Energy}

The appearance in this result that the energies are the product of the masses times $c^2$, the speed of light squared, was the origination of that concept, the famous Einstein's $E = m \cdot c^2$. The concept falls out naturally from applying the Lorentz transforms to the classical definition of kinetic energy. It is somewhat surprising that Einstein was the first to do that inasmuch as it was Lorentz who developed the Lorentz transforms and the Lorentz contractions.

## *Alternative Treatment of the Same Derivation*

If in the above original derivation one proceeds differently from equation *(9)* on, as below, a slightly different result is obtained.

*(9)* $\quad KE = \dfrac{m_r \cdot v^2}{\left[1 - \dfrac{v^2}{c^2}\right]^{\frac{1}{2}}} - m_r \cdot c^2 \cdot \left[1 - \dfrac{v^2}{c^2}\right]^{\frac{1}{2}} - m_r \cdot c^2 \qquad$ [Repeat *(9)* to start]

*(12)* $\quad KE + m_r \cdot c^2 = \dfrac{m_r \cdot v^2}{\left[1 - \dfrac{v^2}{c^2}\right]^{\frac{1}{2}}} - m_r \cdot c^2 \cdot \left[1 - \dfrac{v^2}{c^2}\right]^{\frac{1}{2}} \qquad$ [Move the "$- m_r \cdot c^2$"]

Considering and evaluating the three terms of equation *(12)*:

*(13)* $\quad KE + m_r \cdot c^2 \;$ = Kinetic plus rest energies
$\qquad\qquad\qquad\quad\;$ = Total Energy
$\qquad\qquad\qquad\quad\;$ = $m_v \cdot c^2$

*(14)* $\quad \dfrac{m_r \cdot v^2}{\left[1 - \dfrac{v^2}{c^2}\right]^{\frac{1}{2}}} \;$ = A relativistically increased energy of motion which equals zero when $v = 0$.

$\qquad\qquad\qquad\quad\;$ = $m_v \cdot v^2$

*(15)* $\quad m_r \cdot c^2 \cdot \left[1 - \dfrac{v^2}{c^2}\right]^{\frac{1}{2}} \;$ = A relativistically reduced rest energy which equals the at rest energy when $v = 0$
$\qquad\qquad\qquad\qquad\qquad\;$ = Equation *(13)* − Equation *(14)*
$\qquad\qquad\qquad\qquad\qquad\;$ = $m_v \cdot c^2 - m_v \cdot v^2$

the result is that equation *(12)* is equivalent to

$$(16) \begin{bmatrix} \text{Total} \\ \text{Energy} \end{bmatrix} = \begin{bmatrix} \text{Energy in} \\ \text{Kinetic Form} \end{bmatrix} + \begin{bmatrix} \text{Energy in} \\ \text{Rest Form} \end{bmatrix}$$

$$m_v \cdot c^2 = m_v \cdot v^2 + m_v \cdot (c^2 - v^2)$$

and (dividing the above energy equation by $c^2$ to obtain an equation in mass)

$$(17) \begin{bmatrix} \text{Total} \\ \text{Mass} \end{bmatrix} = \begin{bmatrix} \text{Mass in} \\ \text{Kinetic Form} \end{bmatrix} + \begin{bmatrix} \text{Mass in} \\ \text{Rest Form} \end{bmatrix}$$

$$m_v = m_v \cdot v^2/c^2 + m_v \cdot (1 - v^2/c^2)$$

Why is the formulation for classical *Kinetic Energy* $KE = \frac{1}{2} \cdot m \cdot v^2$ but *Energy in Kinetic Form* is simply $m \cdot v^2$ without the $\frac{1}{2}$? When dealing with quite small velocities ($v$ very small relative to $c$) the excursion of total energy above rest energy and the excursion of energy in rest form below rest energy are both essentially linear. In that case the portion above the rest case is essentially half of the total excursion above and below the rest case. The classical kinetic energy is then half, $\frac{1}{2} \cdot m \cdot v^2$, the total energy in kinetic form, $m \cdot v^2$, for $[v/c]$ quite small.

### *Application to the Problem of the Matter Wave*

Thus the traditional view of kinetic energy as the energy increase due to motion may not be valid as a description of the processes taking place. Before the encountering of the relativistic change in mass with velocity the traditional view did not lead to problems in spite of its apparently being an over-simplification.

Using mass- and energy-in-kinetic-form to obtain the frequency of the matter wave proceeds as follows.

$$(18) \quad f_{mw} = \frac{m_v \cdot v^2}{h} \quad \text{[equation (6), but using } W_v, \text{ energy-in-kinetic-form, for } W_k, \text{ kinetic energy]}$$

Using this result for matter wave frequency and using the same relativistic mass, $m_v$, in equation *(5)* for the matter wavelength the velocity of the matter wave then is

$$(19) \quad v_{mw} = f_{mw} \cdot \lambda_{mw}$$

$$= \left[\frac{m_v \cdot v^2}{h}\right] \cdot \left[\frac{h}{m_v \cdot v}\right]$$

$$= v$$

and the wave is traveling with and as the particle.

On that basis the wave aspect of matter is then established both experimentally (Davison and Germer and their successors) and theoretically (the above development). That gives new significance to the fact, observed at the time of Bohr's development of the relationship between atomic line spectra and atomic orbital structure, that the stable orbits of atomic electrons are an integer multiple of the orbital electron's matter wave length.

The fact of the stable orbits has long been accepted without a specific reason, a specific operative cause, for those orbits and only those orbits being stable. The matter wave of the orbiting electron now provides an operative reason, as follows.

For the orbit to be stable it must be the same for each pass, pass after pass. If each pass includes exactly an integer number of the orbital electron's matter wave lengths then each pass is the same in that regard. But if, for example, the orbital path length contains only $9/10$ of a matter wave length, $9/10$ of the matter wave period, then the next pass will contain the missing $1/10$ of the matter wave length or wave period plus $8/10$ of the next, and so on. The matter wave being sinusoidal in form, the successive orbital passes will be all different.

It is this behavior which operatively causes the "stable orbits", and only those orbits, to be stable. It has nothing to do with angular momentum nor quantization of angular momentum. For the angular momentum hypothesis there is no underlying reason nor mechanism to produce stability or instability. The quantization of angular momentum concept is merely a defined condition, without operative cause, just as were the "stable orbits" it seeks to explain until their here being justified in terms of the operative matter wave behavior

The statement that the orbital electron's angular momentum is quantized, as in the following traditional equation

$$(20) \quad m \cdot v \cdot R = n \cdot \frac{h}{2\pi} \qquad [n = 1, 2, \ldots]$$

is merely a mis-arrangement of

$$(21) \quad 2\pi \cdot R = n \cdot \frac{h}{m \cdot v} = n \cdot \lambda_{mw} \qquad [n = 1, 2, \ldots]$$

a statement that the orbital path length, $2\pi \cdot R$, must be an integral number of matter wavelengths, $n \cdot \lambda_{mw}$, long. The latter statement has a clear, simple, operational reason for its necessity. The former statement is arbitrary and is justified only because it produces the correct result, even if without an underlying rational reason.